# The Evening Tutoring Center at the United States Air Force Academy

Michael Courtney, PhD
United States Air Force Academy
Michael.Courtney@usafa.edu

**Abstract:** The United States Air Force Academy (USAFA) has opened an Evening Tutoring Center to provide after-hours tutoring every evening before class days. The center focuses on first and second year courses in challenging quantitative disciplines: mathematics, physics, chemistry, and engineering mechanics. Staffed exclusively by faculty-level instructors, the center offers extra instruction to all first and second year cadets. Early demand has been remarkable, and early indications are that cadets who visit the center for extra instruction in mathematics perform better than peers with comparable backgrounds.

## I. Introduction

Maintaining historical standards of academic rigor entails unique challenges as competition stiffens for the best students, mathematical prowess of high school graduates declines, intercollegiate athletes face the same core course requirements as other cadets, and technological distractions such as internet gaming, social networking sites, and mobile communication devices challenge both character and work ethic. For several years, Student Academic Services at USAFA has offered coursework in essential study skills and college level reading, as well as a Writing Center to provide tutoring in writing and communication skills. This year, Student Academic Services at USAFA added a Quantitative Reasoning/Evening Tutoring Center to provide additional instruction in the evening (from 1800-2200 hours) in first and second year courses in Mathematics, Physics, Chemistry, and Engineering Mechanics.

The Quantitative Reasoning/Evening Tutoring Center is staffed by four faculty instructors (one in each discipline) who teach one section of a core course in the day and provide extra instruction during the evening hours. These four instructors are highly qualified, with three PhD's and one M.S. and a combined total of 100 years of prior teaching experience. Tutoring instructors new to USAFA participated in the New Instructor Training program in their respective departments, and each is a full participant in departmental matters related to the course they are assigned to teach. The Chemistry tutor is responsible for tutoring both the first year Chemistry sequence as well as second year sequence in Organic Chemistry. The Physics tutor is responsible for tutoring Physics 1 (Mechanics mostly) and Physics 2 (Electricity and Magnetism). The Engineering Mechanics tutor tutors the single, 200 level course in Engineering Mechanics. The Mathematics tutor is responsible for Pre-calculus, Calculus 1, Calculus 2, Calculus 3, and Differential Equations. This article describes the philosophy, process, and early outcomes of tutoring Mathematics at the Evening Tutoring Center, as well as ideas for meeting the challenge of rapidly increasing demand for additional instruction after hours.

## II. Philosophy

The main goal of the service academies is developing officers of character. Quantitative prowess and analytical reasoning contribute to winning wars, but character counts for more. The Dean of Faculty is keen on imparting ample contributions from both Athens (wisdom and knowledge) and Sparta (warrior ethos) to cadets, and due consideration is given to train cadets with the character needed to maintain principles of honor and their commissioning oath to support and defend the Constitution. This is no dichotomy: cadets who aspire to be officers of character must be willing to acquire the wisdom and knowledge necessary for the performance of their duties.

So while the educational philosophy of the tutoring center is customer focused (how can we help the cadet succeed in the topic at hand), the longer term focus is in preparing cadets who are responsible, disciplined, mature, and hard working in the service of our country. Key questions we ask ourselves are "Have I encouraged the cadet to grow in the habits and disciplines of a good officer?" and "Have I encouraged the cadet to take ownership of his assigned work as important preparation for future duties?"

Cadets utilize the Evening Tutoring Center for various reasons: some have scheduling conflicts with daytime extra instruction, some have weaker backgrounds in Mathematics, some have not yet learned the proper habits of mind and disciplines of time management needed for success. Most are still teenagers experiencing new levels of pressure and responsibility. They need encouragement and practical advice to rise to that challenge, along with the reminder that they are capable of tremendous accomplishment with the proper application of time and effort.

The role of the tutor is distinct from the role of classroom instructors. Classroom instructors have a dual burden: 1) they are teachers who impart knowledge and guide cadets as they accomplish the required learning objectives, and 2) they are gatekeepers who must ensure quality of learning by developing assessments that accurately reflect the level of student accomplishment. Classroom instructors bear both the carrot (promises and joys of learning the subject) and the stick (firm requirements and consequences if the cadet fails to meet course requirements). In contrast, tutors bear mostly the promises and joys. The tutor gets to be the "good cop" who sympathizes with realities of cadet life and academic challenges and comes along side to help the cadet meet the requirements. Tutors have the advantage of connecting the last few dots so cadets can see the picture clearly. Tutors get to be the voice that says "you can" when other voices, internal and external, are saying "you can't" or "you won't."

As with everything in life, success is conditional, but the tutor breaks down complex challenges into manageable portions so that cadets can see their way through the maze. Quantitative problem solving is analyzing the problem and planning a strategy more mature than formula roulette (hoping to get lucky by applying formulas without really understanding why a given formula is appropriate in a specific context). Before beginning to take a derivative or integral, one must consider how the derivative or integral is related to the broader problem, and one must consider why one has chosen a given approach from the array of possibilities.

### III. Process

Earlier programs offered by Student Academic Services grew gradually as the word circulated that additional assistance was available in writing, study skills, and college-level reading. Open four hours each evening, Sunday-Thursday, The Evening Tutoring Center's initial plan was to focus on individualized attention in scheduled one-on-one tutoring sessions. In the first week of operation, it became clear that meeting cadet demand with available staff would require a combination of individual and group sessions.

In Mathematics, ten hours of course-specific walk-in sessions are scheduled each week, according to demand projected from consideration of assignments, graded reviews, and historical demand. These walk-in sessions are invariably group sessions with anywhere from 3 (ordinary homework help) to 23 (graded review the next day) cadets in attendance. The remaining 10 evening hours each week are used for individualized instruction for which students sign up using an on-line calendar (SharePoint) on a first-come, first-served basis. When the Evening Tutoring Center opens each Sunday night, nearly all of the individual time slots are already taken for that week.

Individualized tutoring is a favorite activity of tutoring center faculty. One gets to review cadet work and identify weak areas. One gets to know cadets well: where they are from, what they hope to major in, why they came to USAFA, what study habits they have built, and where they need additional encouragement. One gets to ask if they are satisfied with the tutoring services. Most of all, one gets the joy of seeing the light go on, observe cadets grow in confidence and ability over the course of the semester, and have cadets arrive beaming and announcing their recent triumph over the latest graded review.

Tutoring group sessions is a greater challenge. Up to groups of three or four, I can still develop a good sense for where each cadet is getting stuck and allow the cadets to work problems while I offer corrections, advice, and encouragement. Depending on the group, once there are more than 5-6, I shift to modeling good problem solving techniques at the board while keeping students engaged by asking them what the next step is and why. Occasionally, group work among students can be productive under the tutor's guidance, but one must be careful about the "blind leading the blind" and bad problem solving habits being propagated. Cadets love short cuts, and the tutor's job is to model problem solving methodologies that are well-considered, well-communicated, and easy to review for errors if the bottom line doesn't make sense. Having cadets help each other is more productive in second year courses where there is more maturity to seek genuine understanding and less tendency to seek the shortest path to the "right answer."

**IV. Demand**
Demand is dominated by Calculus 1 followed closely by Calculus 3. Demand for other courses is far below demand in Calculus 1 and 3. Reasons for Calculus 1 demand are obvious. Every institution has students struggle in calculus even though they have rigorous admission standards. Calculus 1 is the first college level math course for many cadets, and it requires a level of discipline and good habits of mind that many cadets are still developing. We attribute Calculus 3 demand to the fact that the Calculus 1 and Calculus 2 are designed as three credit hour courses geared for all majors. In contrast, Calculus 3 is comparable to multivariable calculus courses in engineering and science majors at top 20 schools and represents a significant step up in the mathematical maturity, quantitative reasoning, and problem solving skills compared with the first year Calculus sequence. It is harder for cadets to move into our multivariable Calculus course with a background of 6 credit hours in Calculus than it is for students at other institutions with 8-10 hours of pre-requisite Calculus courses.

In the first semester of operation (Fall 2009), the tutoring center served over 2500 cadets, over 800 in Math. On a typical evening, the tutoring center sees 2-4 cadets for individualized sessions in math, and 5-10 cadets for group walk-in sessions. Most students will make good use of the whole hour, if they come to an hour-long walk-in session or schedule an hour long appointment. Students are encouraged to come prepared and most show up with a specific list of homework problems they need help with and/or specific topics to discuss. Students needing less help will typically only schedule a half-hour individual session.

**V. Outcomes and Discussion**
Cadets served by the Evening Tutoring Center tend to have weaker backgrounds than the cadet wing as a whole. For example, the cadets enrolled in Calculus 1 who visited the tutoring center in the first two weeks to prepare for the first Fundamental Skills Exam (FSE), which covers high-school Algebra, averaged 52.1%±2.9% on the Algebra portion of the departmental placement exam, whereas the cohort of cadets currently enrolled in Calculus 1 averaged 60.1%±0.6% on the Algebra portion of the placement exam. (A score of 38% is considered passing, sufficient for placement into Calculus 1. Higher scores are required for placement into Calculus 2 and 3.) However, outcomes of cadets who utilized evening tutoring services tend to be comparable with the cadet wing as a whole. For example, the cadets who visited in preparation for the first FSE scored an average of 76.5%±4.6%, whereas the complete cohort of cadets in Calculus 1 averaged 78.4%±0.6% Consequently, cadets who used our services improved their score by an average of 24.4%±3.6% between the two events, compared with the complete Calculus 1 cohort who improved their average score by 18.3%±0.6%.

This is no surprise. Every math teacher can attest that cadets who receive extra instruction tend to do better than cadets who do not. Individualized attention can identify and address a cadet's specific struggles. The service academies are well-known for instructor availability. However, there is clearly additional need for extra instruction beyond traditional office hours. Intercollegiate athletes have mornings filled with classes and afternoons filled with athletic practice. Evenings are the only time they regularly have available for extra instruction. Earning 147 credit hours in 8 semesters, along with military training and duties, in addition to the required intramural participation also makes for full days for many other cadets. The academic call to quarters each evening provides a block of time when many cadets can make use of available extra instruction. Cadets, instructors, and course directors have all expressed appreciation for the availability of evening tutoring services in challenging quantitative disciplines. (MIT has had evening tutoring available for almost 30 years through their Office of Minority Education. Principally staffed by graduate students, their tutoring services are open to all students. I served as a Calculus tutor there while in graduate school, and their system serves a wide variety of minority and non-minority students.)

As word circulates regarding the availability of the Evening Tutoring Center, we expect demand to continue to increase. Additional increases are expected as cadets already using our services succeed and move on to courses currently under-represented in our demand profile (Calculus 2 and Differential Equations). Maintaining adequate service will prove challenging with the current staffing level. As budgets and staffing permit, we foresee several enhancements to meet this challenge.

The Writing Center already makes use of faculty volunteers to approximately double its tutoring manpower compared with their dedicated tutors alone. The Chemistry department was the first quantitative discipline to send faculty to augment the efforts of the dedicated Chemistry tutor. With the size of the USAFA Department of Mathematical Sciences, evening tutoring availability in Mathematics would be doubled if each department member volunteered for evening tutoring once every two months.

Another idea for better serving students with existing manpower would be to provide on-line availability to high-quality exemplar solutions for assigned homework problems or close analogues. Of course, the answers to many problems are in the back of the book, and answers are often posted on-line to selected homework problems. However, published answers and solutions are of variable quality, often overemphasizing equations rather than modeling sound problem solving. Potential benefits of this approach are limited, particularly for first year students, because at this stage of their mathematical development, many students are not yet sufficiently disciplined to focus and gain full benefit from even the highest quality written solutions. Rather than do their best on a given problem until legitimately stuck, view the model solution to focus on the solution process, and then return to complete their own solution, students rely too heavily on written solutions.

Another option is to provide a list of topic-specific on-line resources. For example, the tutoring center is incapable of providing adequate resources to meet demand when a significant number of cadets miss the same lecture while traveling to an intercollegiate athletic event, and tutoring is not really a substitute for a quality lecture. As part of their Open Courseware materials, MIT publishes a complete set of on-line high-quality video lectures covering nearly every topic in differential and integral calculus, multivariable calculus, and differential equations.[2] While the MIT lecture style differs from the Thayer method favored at USMA and the similar hands-on style favored at USAFA, these lectures can be a valuable supplement for cadets who miss a specific lecture or whose learning style is better served by a traditional lecture format. There are also a set of high-quality Calculus videos published by the University of Houston [3] that are more tutorial in style and may be better suited for students who are building on a classroom lesson.

In addition, a number of the mathematical articles in Wikipedia are of surprisingly high quality (given that anyone can edit them and many editors are anonymous) and can provide an explanation that is complementary to the textbook and classroom presentation. I have vetted a number of these entries and compiled a list of quality entries for students to use at my classroom sub-page of the Calculus 1 SharePoint web site. While existing internet resources can be valuable, I recommend careful vetting for quality as well as compiling an organized list of topics that correlate well with specific material in the course.

Cadet feedback indicates they benefit more from on-line video lectures than from on-line written material. In addition, tutors are often asked to model problem solving strategies for the same homework exercises repeatedly by different cadets. This suggests we can serve more cadets with limited manpower by producing videos that model well-communicated solutions to assigned exercises or close analogues. Cadets would probably prefer that we provide model solutions to assigned exercises, but there is probably more benefit in modeling close analogues and allowing cadets to do the mental labor bridging the gap between the close analogue and the assigned problem. Listing each on-line video tutorial according to textbook analogues would quickly guide cadets to the available assistance when they are stuck on a specific problem or concept.

In summary, early feedback indicates that the availability of evening tutoring will enhance student success in challenging quantitative coursework, particularly by better serving cadets who have scheduling difficulties meeting with their primary course instructors during daytime office hours. Meeting cadet demand for evening tutoring will be an ongoing challenge, but we have identified several avenues that should prove productive as we move forward.

**REFERENCES**

1. A more complete discussion of Fundamental Skills Exams at USAFA is found in: *Teaching Fundamental Skills at the United States Air Force Academy,* James S Rolf, Michael A. Brilleslyper, and Andrew X. Richardson, Mathematica Militaris, Volume 15, Issue 1, Spring 2005.

2. M.I.T. Open Course Ware, http://ocw.mit.edu

3. Selwyn Hollis, Video Calculus, University of Houston, Department of Mathematics, http://www.online.math.uh.edu/HoustonACT/videocalculus/index.html